\documentclass{mem}
\usepackage{natbib}
\usepackage{txfonts}
\usepackage{balance}
\usepackage{graphicx}
\usepackage[a4paper]{hyperref}
\begin{document}

\def\age{\mbox{$Age$}}
\def\rhoDM{\mbox{$\langle \rho_{\rm DM} \rangle$}}
\def\eSF{\mbox{$\epsilon_{\rm SF}$}}
\def\Re{\mbox{$R_{\rm eff}$}}
\def\RE{\mbox{$R_{\rm E}$}}
\def\Msun{\mbox{$M_\odot$}}
\def\Mtot{\mbox{$M_{\rm tot}$}}
\def\Lsun{\mbox{$L_\odot$}}
\def\Ysun{\mbox{$\Upsilon_\odot$}}
\def\ML{\mbox{$M/L$}}
\def\Yst{\mbox{$\Upsilon_{*}$}}
\def\YstB{\mbox{$\Upsilon_{*,B}$}}
\def\YstK{\mbox{$\Upsilon_{*,K}$}}
\def\YBe{\mbox{$\Upsilon_{\rm B,eff}$}}
\def\Ydyn{\mbox{$\Upsilon_{\rm dyn}$}}
\def\Ytot{\mbox{$\Upsilon_{\rm tot}$}}
\def\Yvir{\mbox{$\Upsilon_{\rm vir}$}}
\def\mst{\mbox{$M_{\star}$}}
\def\mpr{\mbox{$M_{\rm p}$}}
\def\LB{\mbox{$L_{\rm B}$}}
\def\Yre{\mbox{$\Upsilon_{B}(\Re)$}}
\def\Ypre{\mbox{$\Upsilon_{B,p}(\Re)$}}
\def\MB{\mbox{$M_{B}$}}
\def\sigS{\mbox{$\sigma_{\rm SIS}$}}
\def\sigN{\mbox{$\sigma_{\rm NIS}$}}
\def\rvir{\mbox{$r_{\rm vir}$}}
\def\CMvir{\mbox{$c_{\rm vir}-M_{\rm vir}$}}
\def\mdm{\mbox{$M_{\rm DM}$}}
\def\Mvir{\mbox{$M_{\rm vir}$}}
\def\Ss{\textsf{S sample}}
\def\Cs{\textsf{C sample}}
\def\fpdm{\mbox{$f_{\rm p, DM}$}}
\def\fdm{\mbox{$f_{\rm DM}$}}
\def\lsim{\mathrel{\rlap{\lower3.5pt\hbox{\hskip0.5pt$\sim$}}
    \raise0.5pt\hbox{$<$}}}                
\def\gsim{~\rlap{$>$}{\lower 1.0ex\hbox{$\sim$}}}

\def\Msun{\mbox{$M_\odot$}}
\def\Zsun{\mbox{$Z_{\odot}$}}
\def\lsim{\mathrel{\rlap{\lower3.5pt\hbox{\hskip0.5pt$\sim$}}
    \raise0.5pt\hbox{$<$}}}
\def\gsim{~\rlap{$>$}{\lower 1.0ex\hbox{$\sim$}}}

\def\tg{\mbox{$t_{\rm gal}$}}
\def\tA{\mbox{$t_{\rm AGN}$}}
\def\tq{\mbox{$t_{\rm q}$}}
\def\sig{\mbox{$\sigma$}}
\def\sigc{\mbox{$\sigma_{0}$}}
\def\Re{\mbox{$R_{\rm eff}$}}
\def\eSF{\mbox{$\epsilon_{\rm SF}$}}
\def\mst{\mbox{$M_{*}$}}

\def\gZ{\mbox{$\nabla_{\rm Z}$}}
\def\gage{\mbox{$\nabla_{\rm age}$}}

\def\gug{\mbox{$\nabla_{\rm u-g}$}}
\def\ggr{\mbox{$\nabla_{\rm g-r}$}}
\def\ggi{\mbox{$\nabla_{\rm g-i}$}}
\def\ggz{\mbox{$\nabla_{\rm g-z}$}}

\title{Colour and stellar population gradients in galaxies }

   \subtitle{}

\author{
C. \,Tortora\inst{1} \and N.R. \, Napolitano\inst{2} \and
V.F.~Cardone$^{3}$ \and M.~Capaccioli$^{4}$ \and P.~Jetzer$^{1}$
\and R.~Molinaro\inst{2}
          }

  \offprints{C. Tortora}

\institute{ Universit$\rm \ddot{a}$t Z$\rm \ddot{u}$rich, Institut
f$\rm \ddot{u}$r Theoretische Physik, Winterthurerstrasse 190,
CH-8057, Z$\rm \ddot{u}$rich, Switzerland
\email{ctortora@physik.uzh.ch} \and INAF -- Osservatorio
Astronomico di Capodimonte, Salita Moiariello, 16, 80131 - Napoli,
Italy \and Dipartimento di Fisica Generale A. Avogadro,
Universit\`a di Torino and Istituto Nazionale di Fisica Nucleare -
Sezione di Torino, Via Pietro Giuria 1, 10125 - Torino, Italy \and
Dipartimento di Scienze Fisiche, Universit\`{a} di Napoli Federico
II, Compl. Univ. Monte S. Angelo, 80126 - Napoli, Italy}

\authorrunning{Tortora et al.}

\titlerunning{Gradients in local galaxies}

\abstract{We discuss the colour, age and metallicity gradients in
a wide sample of local SDSS early- and late-type galaxies. From
the fitting of stellar population models we find that metallicity
is the main driver of colour gradients and the age in the central
regions is a dominant parameter which rules the scatter in both
metallicity and age gradients. We find a consistency with
independent observations and a set of simulations. From the
comparison with simulations and theoretical considerations we are
able to depict a general picture of a formation scenario.
 }

\maketitle{}

\section{Introduction}\label{sec:intro}

Different physical processes might rule the galaxies' properties
at the global galaxy scale, or act at sub-galactic scales (e.g.
the nuclear regions vs outskirts), such that they are expected to
introduce a gradient of the main stellar properties with the
radius that shall leave observational signatures in galaxy
colours. In fact, color gradients (CGs) are efficient markers of
the stellar properties variations within galaxies, in particular
as they mirror the gradients of star ages and metallicities
(\citealt{Tortora+10CG}, T+10 hereafter). CGs are primarily a tool
to discriminate the two broad formation scenarios (monolithic vs
hierarchical), but more importantly they provide a deeper insight
on the different mechanisms ruling the galaxy evolution. As these
mechanisms depend on the galaxy mass scale, the widest mass (and
luminosity) observational baseline is needed to remark the
relative effectiveness of the different physical processes (such
as merging, AGN, supernovae, stellar feedback, etc) and their
correlation with the observed population gradients.

We will discuss the colour, age and metallicity gradients for a
huge sample of local SDSS early- and late-type galaxies (ETGs and
LTGs, hereafter) from \citet[B+05 hereafter]{Blanton05}, and
analyze the trends with mass. The observed trends are interpreted
by means of quite different physical phenomena at the various mass
scales. See T+10 for further details\footnote{We have used the
structural parameters given by B+05 to derive the color profile
$(X-Y)(R)$ of each galaxy as the differences between the
(logarithmic) surface brightness measurements in the two bands,
$X$ and $Y$. The CG is defined as the angular coefficient of the
relation $X-Y$ vs $\log R/\Re$, where \Re\ is the effective
radius.}.

We assume a set of ``single burst'' synthetic stellar models from
\citet{BC03}, with age and metallicity ($Z$) free to vary. From
the fitting to the observed colours we estimate the galaxy age,
$Z$, and the stellar mass\footnote{We define the age and
metallicity gradients as $\gage =\log [\rm age_{2}/\rm age_{1}]$
and $\gZ=\log [\rm Z_{2}/ \rm Z_{1}]$, where $(\rm age_{i}, \rm
Z_{i})$ with $i=1,2$ are the estimated age and metallicity at $0.1
\Re$ and $\Re$, respectively.}.

\section{Colour gradients}

\begin{figure}[t]
\includegraphics[width=0.24\textwidth,clip]{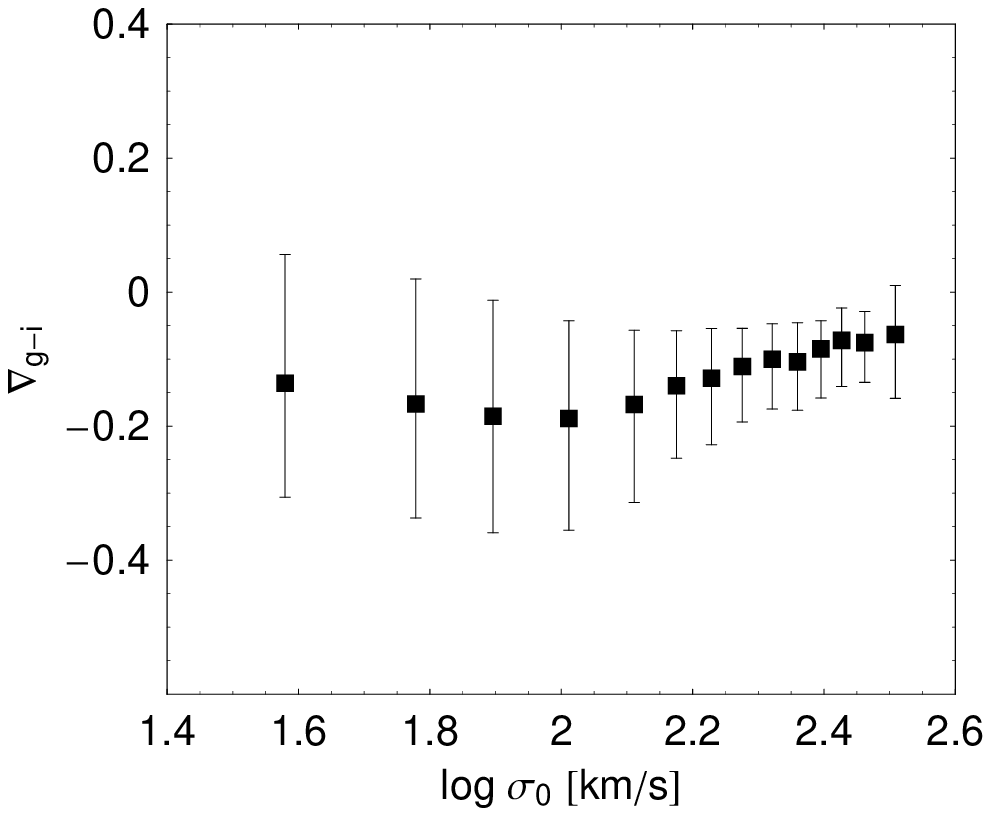}
\includegraphics[width=0.24\textwidth,clip]{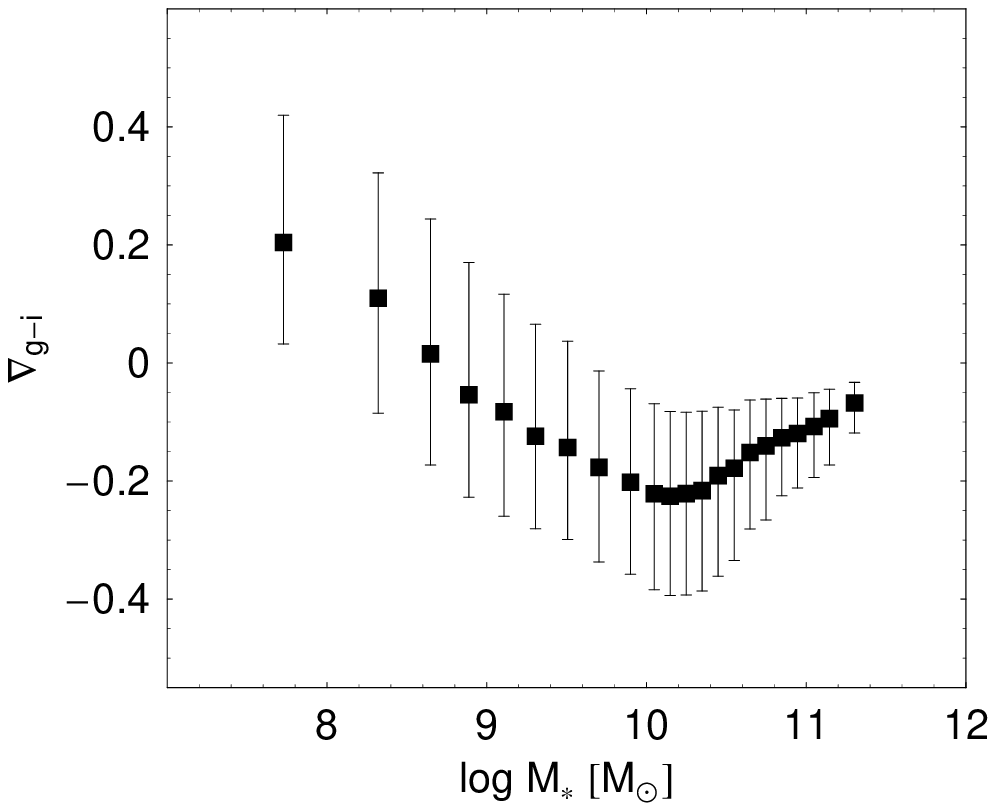}
\caption{\ggi\ as a function of central velocity dispersion (left
panel) and stellar mass (right panel) for the whole sample of
galaxies. Medians and 25-75th quantiles of the sample distribution
in each mass bin are shown.}\label{fig: fig1}
\end{figure}

We start from showing in Fig. \ref{fig: fig1} the results for the
$g-i$ CGs as a function of stellar mass and velocity dispersion
for the whole galaxy sample. The gradients as a function of
velocity dispersion look negative almost everywhere. Although our
sample does not include very massive galaxies ($\mst \gsim
10^{11.5}\, \rm \Msun$), our results seem to show a stable trend
at the large mass scales, pointing to even shallower gradients. We
note here that the smaller range shown by $\ggi$ as a function of
\sigc\ in the left panel of Fig. \ref{fig: fig1} is mainly due to
the mix of the ETGs and LTGs. While, the correlation with stellar
mass show that, starting from the massive/bright end, the CGs
become steeper, with the steepest negative gradients corresponding
to $\mst \sim 10^{10.3}\, \rm \Msun$. From there, the gradients
invert the trend with luminosity and mass, becoming positive at
$\mst \sim 10^{8.7}\, \rm \Msun$.

\begin{figure}[t]
\includegraphics[width=0.24\textwidth,clip]{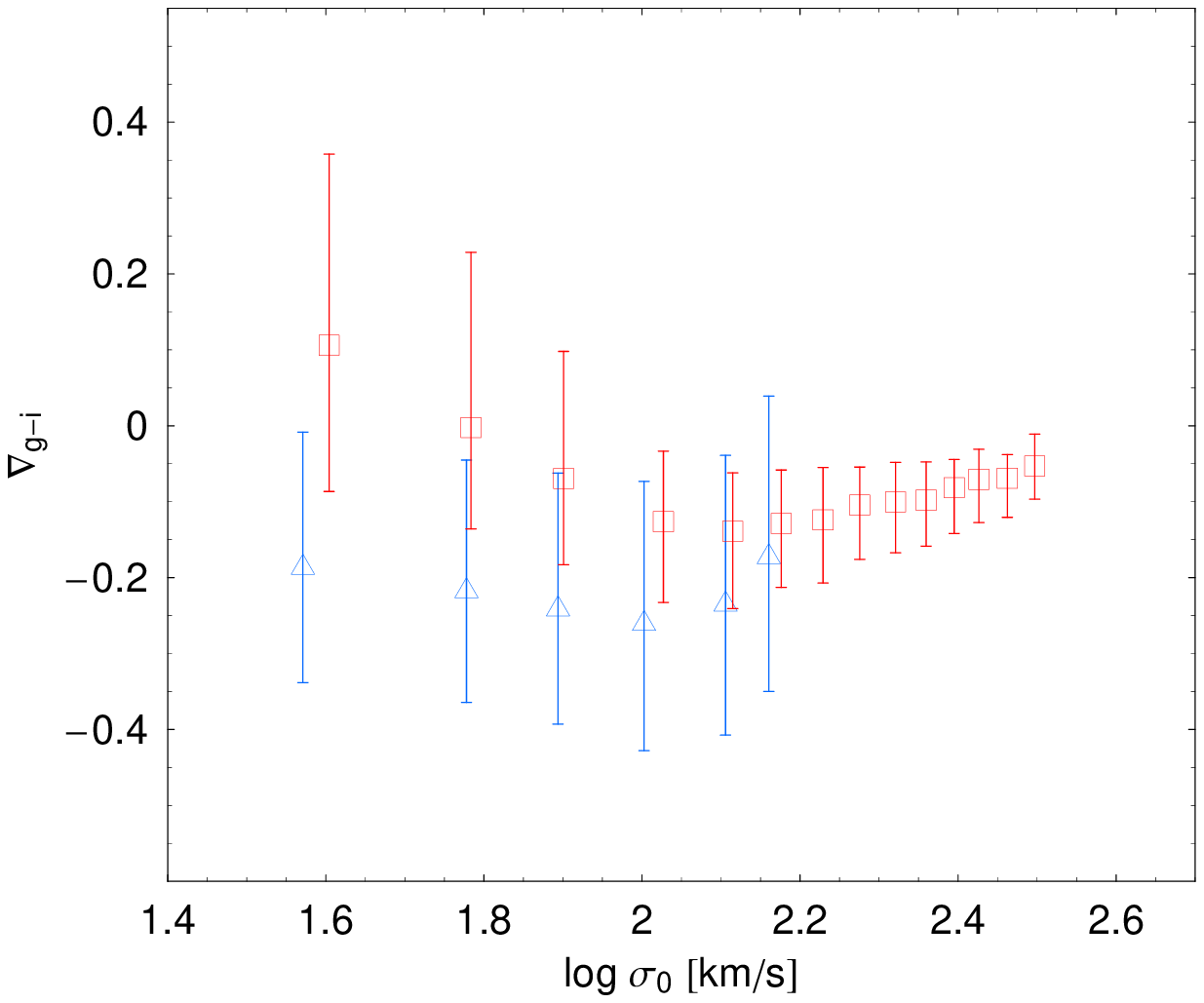}
\includegraphics[width=0.24\textwidth,clip]{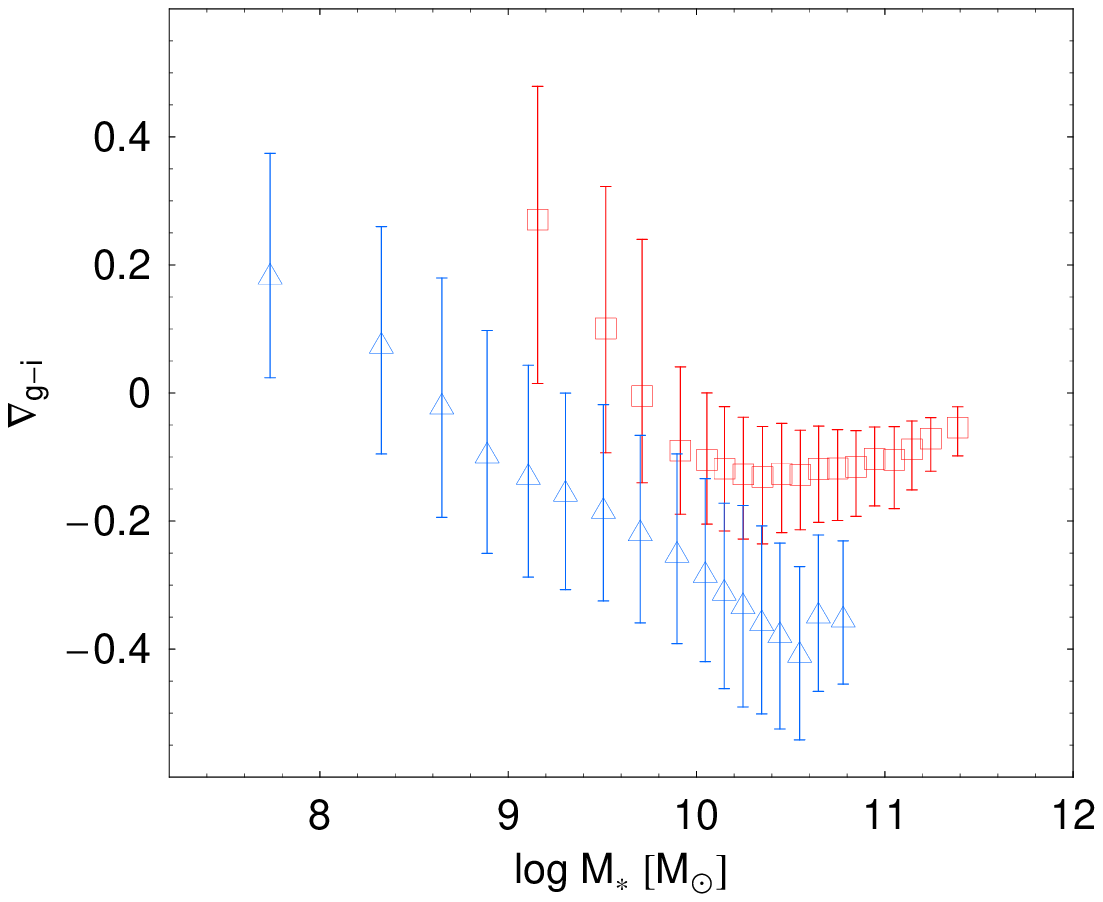}
\caption{\ggi\ as a function of central velocity dispersion (left
panel) and stellar mass (right panel) for ETGs (red symbols) and
LTGs (blue symbols).}\label{fig: fig2}
\end{figure}

From Fig. \ref{fig: fig2}, we see that LTGs show a monotonic
decreasing trend, while a U\,-\,shaped function is found for ETGs
with the gradients definitely decreasing with the mass for $\mst
\lsim 10^{10.3-10.5}\, \rm \Msun$, and a mildly increasing for
larger mass values.  This mass scale is roughly compatible with
the typical luminosity (and mass) scale for ETG dichotomy in the
galaxy structural properties (\citealt{Capaccioli92}) or for
star-forming and passive systems (\citealt{Kauffmann2003}).
Nevertheless, for a fixed stellar mass we observe that ETGs
gradients are, on average, shallower than LTG ones. The same
two\,-\,fold trend is shown for ETGs gradients as function of the
velocity dispersion, while no trend is observed for LTGs
gradients.

\section{Age and metallicity gradients}

\begin{figure}[t]
\includegraphics[width=0.23\textwidth,clip]{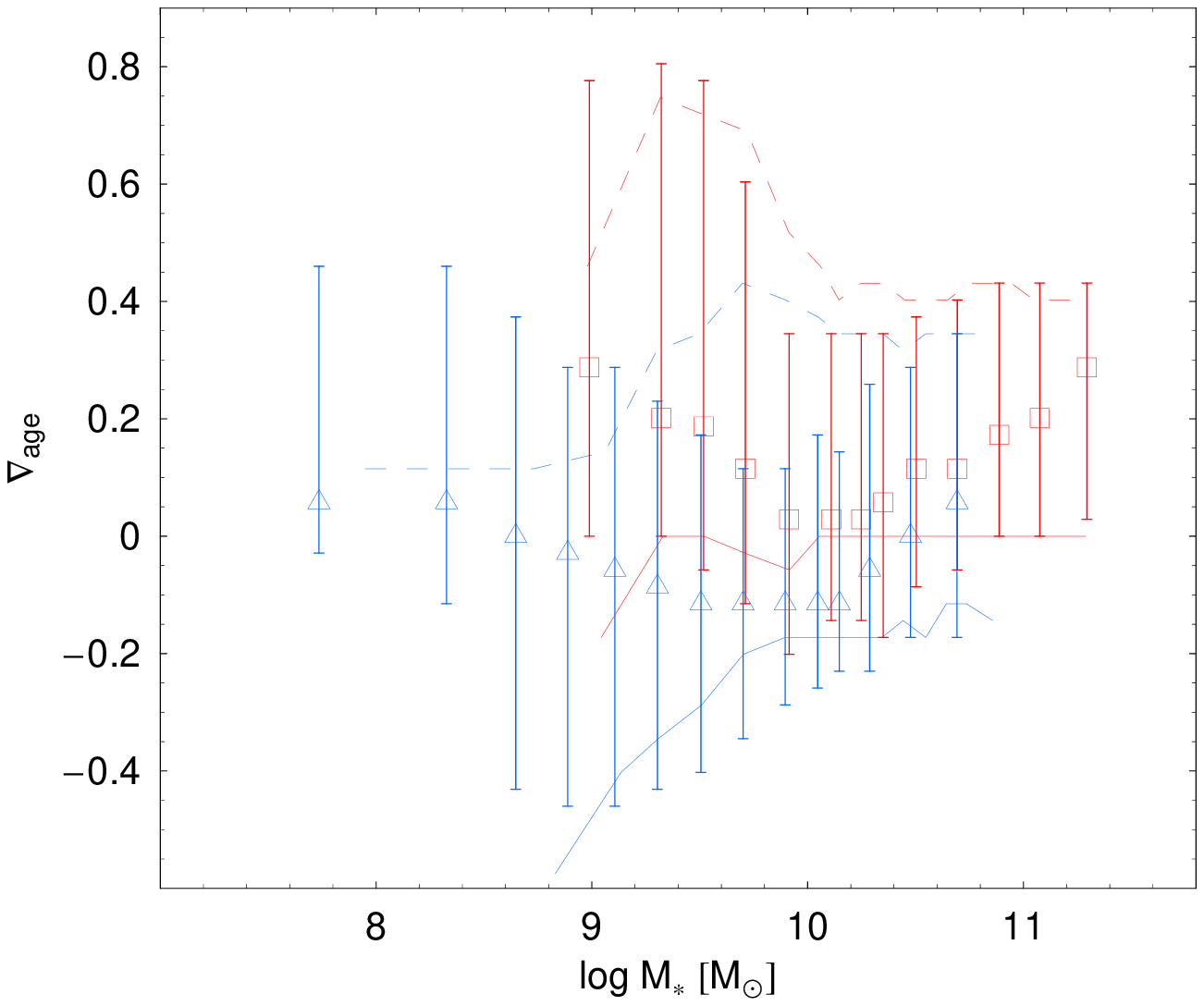}
\includegraphics[width=0.23\textwidth,clip]{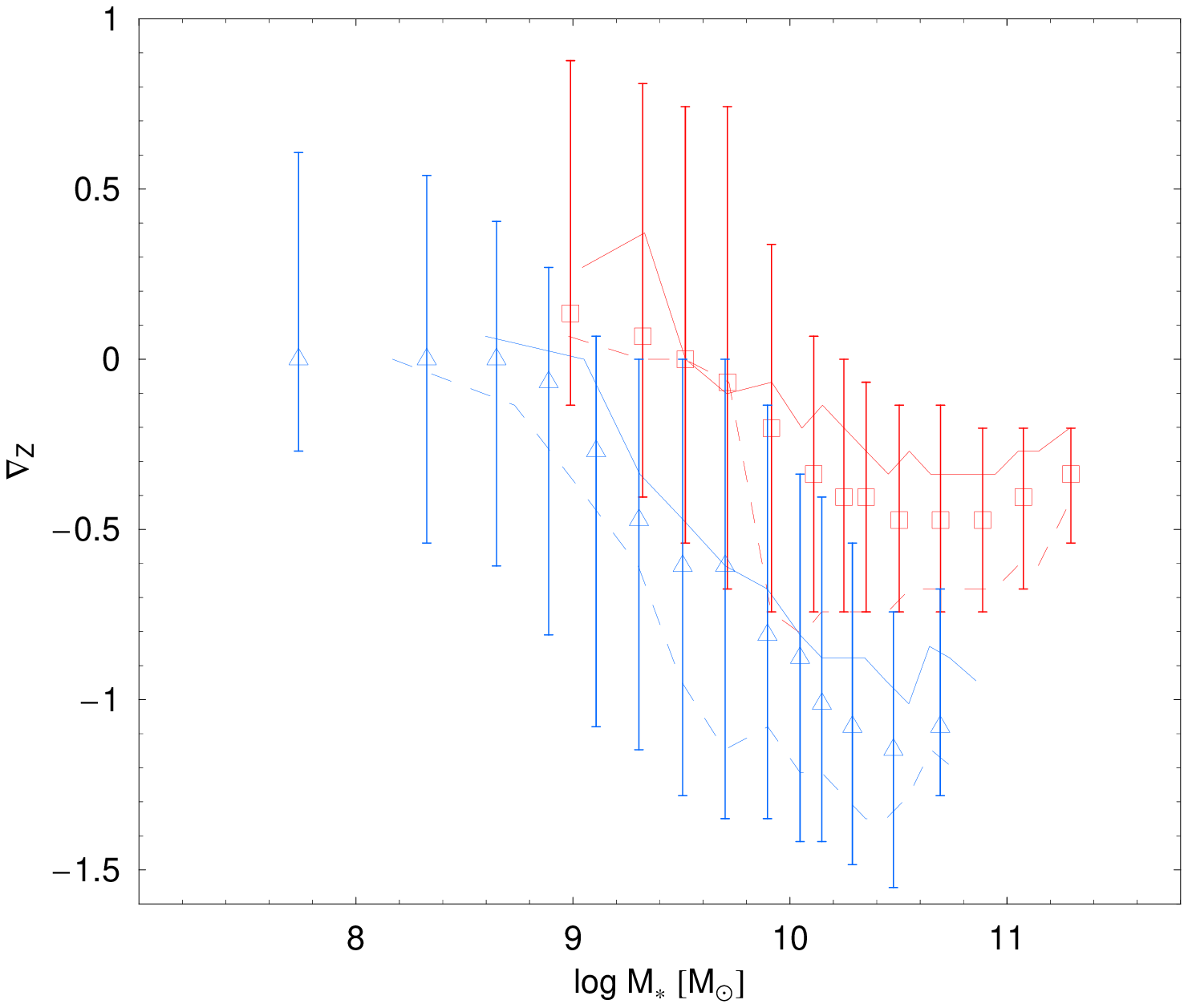}
\caption{Age (left) and metallicity (right) gradients as a
function of stellar mass. The symbols are as in Fig. \ref{fig:
fig2}. Solid and dashed lines are for galaxies with central
stellar populations older and younger than $\rm 6 \, Gyr$,
respectively.}\label{fig: fig3}
\end{figure}
As shown in Fig. \ref{fig: fig3}, for LTGs the \gage\ is about
zero both with mass (and with \sigc, see T+10), strongly
suggesting that CGs should not depend on age gradients of the
galaxy stellar population. Actually, \gZ\ is instead strongly
dependent on \mst, with the lowest metallicity gradients ($\sim
-1$) at the largest masses. ETG age gradients seem basically
featureless when plotted against \sigc\, where we find $\gage > 0$
with typical median values of $\gage \sim 0.2$. On the other hand,
the metallicity gradients show some features with \gZ\ decreasing
for $\log \sigc \lsim 2.2 \, \rm km/s$, and increasing for larger
velocity dispersion, reaching the shallowest values ($\sim -0.2$)
at $\log \sigc \gsim 2.4 \, \rm km/s$ and at the very low $\sigc$
(i.e. $\log \sigc<1.8\, \rm km/s$) where \gZ$\sim0$. This peculiar
trend is mirrored by a similar dependence on the stellar mass.
This two-fold behaviour is also significant when we plot \gage\ as
a function of the stellar mass: \gage\ decreases at $\log \mst
\lsim 10.3-10.5$, and increases in more massive systems. We have
also found that \gage\ and \gZ\ strongly depend on galaxy age: in
particular, if we separate the sample in systems with older and
younger than $6 \, \rm Gyr$ central stars, we obtain different age
and metallicity gradient trends as shown in Fig. \ref{fig: fig3}.
Older systems have gradients which are shallower than younger (in
particular for ETGs) and bracket the average trend of the whole
samples.

\begin{figure*}[t]
\hspace{0.8cm}\includegraphics[width=0.43\textwidth,clip]{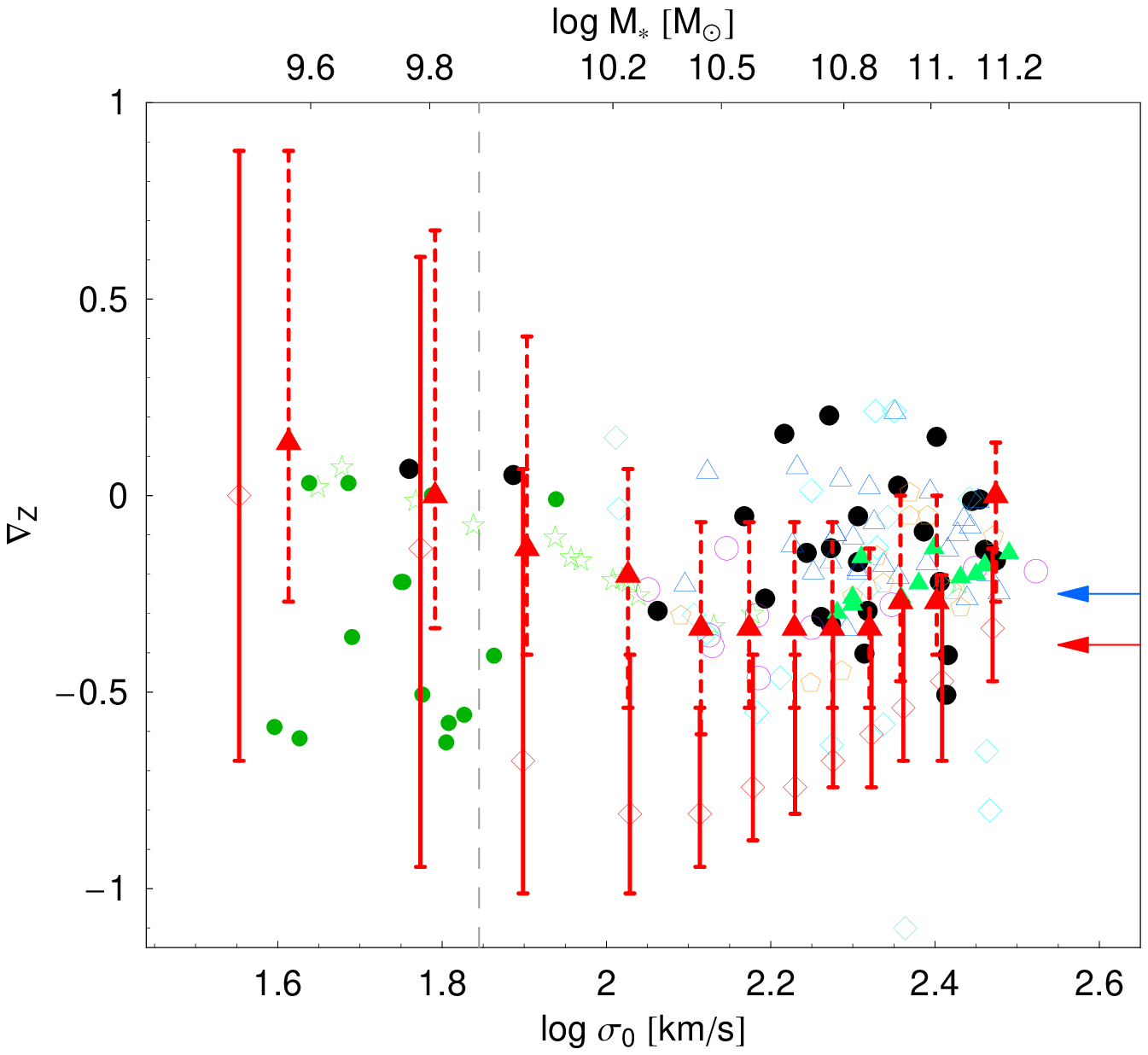}
\includegraphics[width=0.43\textwidth,clip]{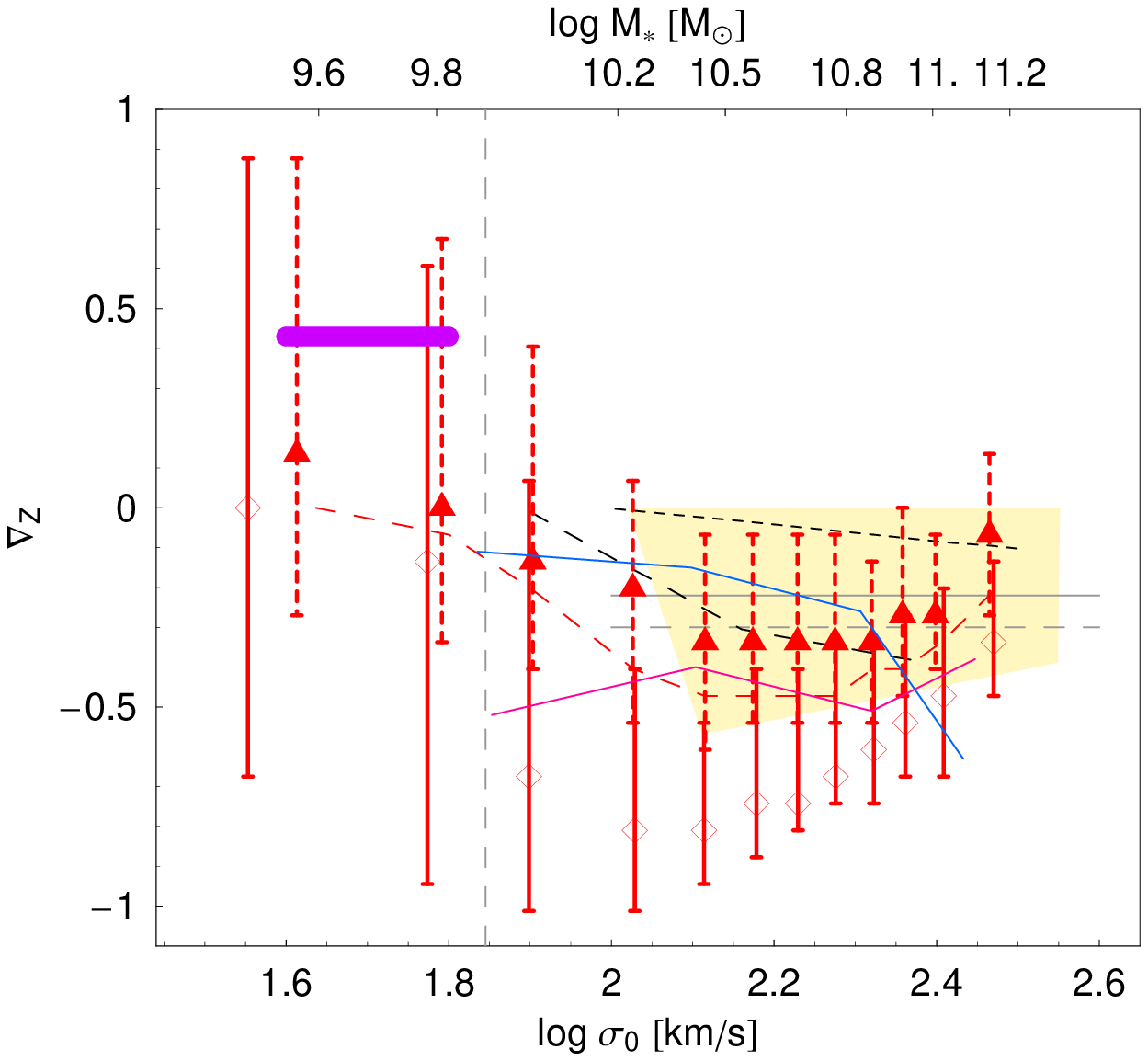}
\caption{{\it Left.} Metallicity gradients versus a set of
observations (see T+10 for further details). {\it Right.}
Metallicity gradients versus simulations. The predictions from
merging models in \citet{BS99} are shown as short dashed line,
dissipative collapse models in \citet{KG03} as long dashed line,
the remnants of major-mergings between gas-rich disk galaxies in
\citet{Hopkins+09a} as yellow shaded region, the typical gradient
for major (continue gray line) and non-major (dashed gray line)
mergings in \citet{Ko04}. The violet thick line shows the
predictions as in the simulation of \citet{Mori+97}; blue and pink
lines are the result from the chemo-dynamical model in
\citet{Kawata01}, respectively for strong and weak supernovae
feedback models. Triangles with dashed bars and boxes with
continue bars are for galaxies centrally older and younger than
$6\, \rm Gyr$; in the right panel the dashed line is the median
trend for all the ETG sample.}\label{fig: fig4}
\end{figure*}

In the left panel of Fig. \ref{fig: fig4} we proceed to a more
detailed comparison of our findings with a set of literature works
which make use of a more sophisticated analysis, although usually
associated to smaller samples (e.g. \citealt{Rawle+09},
\citealt{Spolaor09}). In particular, we concentrate on the ETG
sample. When considering objects with $age_{1}> 6 \, \rm Gyr$, the
agreement with the other studies (generally dealing with old
systems) is remarkably good.

\section{Discussion and conclusions}

We have found some peculiar trends for CGs as a function of
stellar mass, which are mainly driven by metallicity gradients,
although age gradients have a not negligible role. ETGs, have
gradients which are null at $\log \mst \sim 9.5$ and become
steeper at $\log \mst \sim 10.5$, where the trend get inverted and
tend to be shallower at very large masses. LTGs have null colour
and metallicity gradients at $\log \mst \sim 8$, which steepen
with stellar mass and at fixed mass metallicity gradients are
steeper than the ones in ETGs. The scatter in gradients is driven
by the central galaxy age, since centrally older (younger) systems
are found to have shallower (steeper) metallicity gradients.

From the comparison with a set of simulations, which rely on
different physical phenomena (e.g., see for ETGs right panel in
Fig. \ref{fig: fig4}), we are able to point out the main physical
phenomena at different mass scales. Our results seem to support
the idea that the metallicity trend versus the stellar mass for
LTGs is mainly driven by the interplay of gas inflow and winds
from supernovae and evolved stars (\citealt{Ko04}). These
processes tend to increase the central metallicity and prevent the
enrichment of the outer regions. Therefore more massive systems
have on average larger central metallicities which correspond to
steeper negative gradients. Low mass ETGs show the same
correlation (see also the results from the simulation in
\citealt{Tortora+10SIM}), which suggest that these systems might
experience similar phenomena as LTGs (see the comparison with the
different supernovae feedback recipes in Fig. \ref{fig: fig4}).
Moreover, the absolute value of \gZ\ of ETGs is lower than for the
LTGs, probably as a consequence of the dilution of the gradient
due to the higher-density environment where ETGs generally are
located. At very low masses ($\mst \lsim 10^{9.5} \, \rm \Msun$
and $\log \sig_c \lsim 1.8\, \rm km/s$) the \gZ\ turns to even
positive values, which are compatible with the expanding shell
model from \citet{Mori+97}. At larger masses, the shallow
metallicity gradients of ETGs suggest that these have experienced
merging (at a rate that could increase as a function of the
stellar mass) which have diluted the \gZ. Such events might have
taken place in earlier phases of the galaxy evolution, as
indicated by the presence of null age gradients in old systems
(not shown here, but see \citealt{BS99}, \citealt{Ko04},
\citealt{Hopkins+09a}). While, systems with younger cores are
expected to show positive gradients (as found in
\cite{Hopkins+09a}). After the initial gas rich-merging events
that might produce both a larger central metallicity and a
positive age gradient (\citealt{Ko04}), subsequent gas
poor-merging may dilute the positive age gradient with time as
well as make the metallicity gradients to flatten out
(\citealt{Hopkins+09a}). However, a further mechanism that may act
to produce the shallower (or almost null) color and metallicity
gradients in old massive ETGs, with $\mst \gsim 10^{11} \, \rm
\Msun$ and $\log \sigc \gsim 2.4$ (Fig. \ref{fig: fig4}), might be
due to the strong quasar feedback at high redshift
(\citealt{Tortora2009AGN}), while steeper metallicity gradients at
lower masses could be linked to less efficient AGNs.

\begin{acknowledgements}
CT is supported by the Swiss National Science Foundation.
\end{acknowledgements}

\bibliographystyle{aa}

\end{document}